# Ab initio study of exciton insulator phase: Emergent $p$-wave spin textures from spontaneous excitonic condensation

Fang Zhang[1,2], Jiawei Ruan[1,2], Gurjyot Sethi[1,2], Chen Hu[1,2], and Steven G. Louie[1, 2, *]

[1]Department of Physics, University of California at Berkeley, Berkeley, CA, USA.
[2]Materials Sciences Division, Lawrence Berkeley National Laboratory, Berkeley, CA, USA.
*To whom the correspondence should be addressed: sglouie@berkeley.edu

**Abstract:**

An excitonic insulator[1,2] (EI) is a correlated many-body state of electron-hole pairs, potentially leading to high-temperature condensate and superfluidity[3-7]. Despite ever-growing experiments suggesting possible EI states in various materials, direct proofs remain elusive and debated. Here we address the problem by introducing an ab initio methodology, enabling the parameter-free determination of electron-hole pairing order parameter and single-particle excitations within a Bardeen-Cooper-Schrieffer (BCS)-type formalism. Our calculations on monolayer $1T'$-$MoS_2$[8,9] reveals that it is an unconventional EI with a transition temperature $\sim 900K$, breaking spontaneously the crystal's inversion, rotation, and mirror symmetries, while maintaining odd parity and unitarity. We identify several telltale spectroscopic signatures emergent in this EI phase that distinguish it from the band insulator (BI) phase, exemplified with a giant k-dependent $p$-wave spin texture.

Spontaneous symmetry breaking in interacting quantum many-body systems is a fundamental concept underlying the formation of long-range order. A prime example of such a phenomenon is superconductivity. In a superconductor, the condensation of Cooper pairs leads to the emergence of an order parameter, breaking the global $U(1)$ gauge symmetry[10] associated with charge conservation (see Supplementary Materials section 1 '$U(1)$ symmetry breaking'). This order parameter, which may be regarded as the macroscopic wavefunction, describes the superconducting state. The Bogoliubov quasiparticles (BQPs) in the superconducting state, which are coherent superpositions of particle-like and hole-like excitations of the normal state, typically exhibit an energy gap (Fig. 1A). Theoretically, the ground state of an EI phase is characterized by the spontaneous condensation of bound electron-hole pairs, or excitons (Fig. 1B). This condensation breaks a particular local $U(1)$ gauge symmetry associated with a specific operation that is different on the valence and conduction band states[11-13], which we refer to as the $U_X(1)$ symmetry following the terminology used in Ref. 12 (see Supplementary Materials section 2 '$U_X(1)$ symmetry breaking'). Although exciton condensates have been experimentally realized in specially engineered systems, such as bilayer devices with spatially separated electrons and holes[3-6,14,15], the discovery of a material that naturally exhibits excitonic condensation remains elusive.

In superconductors, the breaking of $U(1)$ symmetry gives rise to some unmistakable phenomena such as zero electrical resistance, the Meissner effect, as well as typically the opening of a gap in the quasiparticle spectrum of a metal. In contrast, in terms of quasiparticle excitations, the breaking of $U_X(1)$ symmetry in the EI phase yields another insulator, resulting in quasiparticle wavefunctions that are hybridizations of the original conduction and valence band orbitals of the BI phase (note that $U(1)$ symmetry is preserved in the EI phase). But this kind of hybridization is also a common feature in materials, e.g., systems with band gaps caused by band crossings at the Fermi level and strong spin-orbit coupling (SOC), making it challenging to establish definitive spectroscopic experimental criteria for identifying EIs. Other symmetries of a crystal, however, may be broken upon transition to the EI phase, leading to important telltale signatures of its formation.

Early investigations into possible EIs focused on materials exhibiting charge density waves (CDWs)[16-18] or structural phase transitions[19-22], since condensed electron-hole pairs with finite center-of-mass wavevector $\mathbf{Q}$ may introduce new periodicities in the system. However, these phenomena may also arise from alternative mechanisms, such as anisotropic electron-phonon coupling that induces CDWs, as observed in $TaSe_2$[23]. Recent attention has shifted to monolayer transition metal dichalcogenides (TMDs)[24-29], particularly those in the $1T'$ structure, such as $MoS_2$[24] and $WTe_2$[25,26]. These materials are promising candidates for EI phases due to their low carrier densities and reduced dimensionality, which enhance electron-hole interactions. The absence of CDWs or structure distortions in monolayer $1T'$-TMDs[26,30,31] suggests that, if exciton condensation occurs, it likely involves $\mathbf{Q} = 0$ excitons. Experimental evidences for possible EI states in $1T'$-TMDs primarily arise from considering excitation energies[25,26], as excitation of BQPs (single-particle excitations of the EI phase) would open band gaps in semimetals or enlarge them in insulators. Yet, these observations are not definitive, as similar band gap changes may also result from other mechanisms[32].

Ab initio calculations have played a central role in the past decades in advancing our fundamental understanding of many systems across various subfields; however, they have not been fully applied to EIs due to the absence of a parameter-free ab initio formalism for real materials. Revealing the microscopic nature of EI states from first principles poses two key challenges: 1) accurately describing the electron-hole interaction kernel with the complex many-electron interaction and screening effects in a real material, and 2) solving the typically multi-band BCS-like gap equation to obtain the pairing order parameter and BQP excitations. The ab initio $GW$ plus Bethe-Salpeter equation ($GW$-BSE) approach, which is based on many-body perturbation theory, combines the $GW$ self-energy (where $G$ and $W$ denote the one-particle Green's function and the screened Coulomb interaction, respectively) with the resulting electron-hole interaction kernel to solve the BSE[33,34] to describe exciton states. This method has proven accurate in describing excitonic phenomena across various materials and has been applied to study the unusually large binding energies of excitons in some EI candidates[35,36]. In our $GW$ and $GW$-BSE calculations of the BI phase of monolayer $1T'$-MoS$_2$, the exciton binding energy is found to be 0.28 eV, exceeding the direct band gap of 0.1 eV by nearly a factor of 3, indicative of a strong instability toward an EI state. However, this kind of ab initio studies does not address some critical aspects of EIs, such as the microscopic pairing mechanism, order parameter, or BQP excitations. Conversely, solving the EI gap equations using model Hamiltonians[37,38] may not capture the complex band structure (since often multiple bands are involved) and dielectric screening effects in semimetals or semiconductors, which are important for correctly describing excitonic correlations, and thus less predictive.

In this work, we develop and employ a parameter-free ab initio approach to establish the nature of the EI state and its temperature dependence in monolayer $1T'$-MoS$_2$, arising from the condensation of $\mathbf{Q} = 0$ excitons. By solving the gap equations[1] with the full band structure and electron-hole interaction kernel obtained from ab initio $GW$ and $GW$-BSE calculations, we demonstrate that the ground state of monolayer $1T'$-MoS$_2$ is indeed an EI phase. Our calculations reveal that the spontaneous condensation of excitons (across all spin and band channels) breaks not only the $U_X(1)$ symmetry but also all the point group symmetries of the crystal. This behaviour classifies the EI as an unconventional type, following the terminology used in superconductivity. Additionally, the predicted EI phase exhibits an odd parity in its order parameter (a matrix in band indices and akin to the form of $p$-wave superconductivity), in contrast to an even parity associated with inversion symmetry, and forms a unitary state (a concept analogous to that in superconductivity, as defined later). These microscopic characteristics of the EI phase result in telltale spectroscopic signatures observable in experiments as this system transitions from a BI to an EI. For example, the low-energy BQP excitations in $\mathbf{k}$ space at momenta connected by point group symmetries can differ in energy by up to 4 meV at zero temperature; and the local density of states (LDOS) maps at certain energies exhibit a non-symmetric charge distribution in real space.

More importantly, there is a dramatic change in the electron spin texture of the quasiparticle states between the two phases. The system is nonmagnetic in the BI phase, with the electron spin moments of each degenerate band complex equal to zero at every $\mathbf{k}$ point due to Kramers spin degeneracy. However, in the EI phase, although the magnetization in real space still vanishes everywhere, the BQP states display intriguing, nonzero $\mathbf{k}$-dependent spin textures for all the electron spin components ($S_x$, $S_y$ and $S_z$) in the doubly degenerate valence and conduction bands nearest to the band gap in the Brillouin zone (BZ). In particular, the $\mathbf{k}$-dependent $S_y$ reaches a large value (near 1 $\mu_B$) with an almost perfect $p$-wave symmetry along the $k_y$ direction.

**$GW$-BSE plus BCS self-consistent field method for EIs**

Following Jérome, Rice and Kohn[1], we adopt a reduced multi-band Hamiltonian $\widehat{H}$ for our study of EIs. This Hamiltonian includes two-body electron-hole interactions, given by:

$$\hat{H} = \sum_{v\mathbf{k}} \varepsilon_v(\mathbf{k})\hat{a}^\dagger_{v\mathbf{k}}\hat{a}_{v\mathbf{k}} + \sum_{c\mathbf{k}} \varepsilon_c(\mathbf{k})\hat{b}^\dagger_{c\mathbf{k}}\hat{b}_{c\mathbf{k}} + \sum_{vc\mathbf{k}v'c'\mathbf{k}'} K_{vc\mathbf{k}v'c'\mathbf{k}'}\hat{a}^\dagger_{v\mathbf{k}}\hat{b}_{c\mathbf{k}}\hat{b}^\dagger_{c'\mathbf{k}'}\hat{a}_{v'\mathbf{k}'} \qquad (1)$$

where $\hat{a}_{v\mathbf{k}}$ ($\hat{b}_{c\mathbf{k}}$) annihilates an electron with momentum $\mathbf{k}$ in the $v$-th valence band ($c$-th conduction band) with corresponding band energy $\varepsilon_v(\mathbf{k})$ ($\varepsilon_c(\mathbf{k})$), and $\hat{a}^\dagger_{v\mathbf{k}}$ ($\hat{b}^\dagger_{c\mathbf{k}}$) is its Hermitian conjugate. The term $K_{vc\mathbf{k}v'c'\mathbf{k}'}$ denotes the matrix element $\langle\phi_{v\mathbf{k}}\phi_{c\mathbf{k}}|\hat{K}|\phi_{v'\mathbf{k}'}\phi_{c'\mathbf{k}'}\rangle$, where $\phi_{v\mathbf{k}}$ ($\phi_{c\mathbf{k}}$) are two-component spinor wavefunctions of the $v$-th valence ($c$-th conduction) state with momentum $\mathbf{k}$, and $\hat{K}$ is the electron-hole interaction kernel. We take the ansatz that $\hat{K}$ includes both direct and exchange interactions[7] as in the BSE for excitons within the $GW$-BSE approach[34]. In equation (1), we focus in this study on the subspace of electron-hole pairs with total momentum $\mathbf{Q} = 0$, as no evidence for finite momentum pairing (which would induce CDWs) was observed in Raman spectra[26] or tunneling microscopy[30,31] of monolayer $1T'$-TMDs. Accurate ab initio input for the band energies $\varepsilon_n(\mathbf{k})$ and the electron-hole interaction kernel $\hat{K}$ is essential for this approach. We use the ab initio $GW$ and $GW$-BSE methods[33,34] to obtain these ingredients (see Supplementary Materials section 3 '*GW* and *GW*-BSE calculations of electron-hole kernel and excitons' and fig. S1), with careful treatment of Coulomb interaction and screening in atomically thin two-dimensional (2D) materials[39].

In general, solving the Hamiltonian $\hat{H}$ in equation (1) represents a quantum many-body problem that is challenging to solve exactly. However, it may be treated by a mean-field approach self-consistently, which has proven successful in the study of superconductors within the BCS framework[10]. In fact, the mean-field approximation is expected to perform even better here than in BCS superconductors, as the neglected local quantum fluctuations should have little impact on the long-range character of excitonic correlations[40]. We define a mean-field Hamiltonian $\hat{H}_{MF}$, in which the two-body interaction term in $\hat{H}$ is approximated by a one-body term interacting with a mean field $\Delta_{vc\mathbf{k}}$, which is the order parameter (a matrix in band indices at each $\mathbf{k}$ point) for the EI phase:

$$\Delta_{vc\mathbf{k}} = \sum_{v'c'\mathbf{k}'} K_{v'c'\mathbf{k}'vc\mathbf{k}} \langle \hat{a}^\dagger_{v'\mathbf{k}'} \hat{b}_{c'\mathbf{k}'} \rangle \qquad (2)$$

where $\langle \hat{O} \rangle$ (with $\hat{O} = \hat{a}^\dagger_{v'\mathbf{k}'}\hat{b}_{c'\mathbf{k}'}$ here) denotes the value $\mathrm{tr}[e^{-\hat{H}_{MF}/k_BT}\hat{O}]/\mathrm{tr}[e^{-\hat{H}_{MF}/k_BT}]$ (with $k_B$ the Boltzmann constant and $T$ the temperature). The mean-field Hamiltonian $\hat{H}_{MF}$ is diagonalized using a linear Bogoliubov transformation[41,42] $\hat{B}$. The diagonal elements of $\hat{B}^\dagger\hat{H}_{MF}\hat{B}$ correspond to the BQP excitation energies $E_n(\mathbf{k})$ in the EI phase, and the associated BQP wavefunctions are denoted $|\psi_{n\mathbf{k}}\rangle$. The transformation $\hat{B}$ and the order parameter $\Delta_{vc\mathbf{k}}$ are coupled through a set of self-consistent, nonlinear equations, which we solved iteratively. The final solutions for $\hat{B}$ and $\Delta_{vc\mathbf{k}}$ are very stable upon convergence (fig. S2). The ab initio framework developed here, combining the $GW$-BSE method for the electron-hole kernel and self-consistent field (SCF) method for solving the gap equations (the $GW$-BSE-SCF method in short; see Supplementary Materials section 4 'GW-BSE-SCF method for EI phase'), offers a general approach that can be extended to study EI phases in various intrinsic materials, bilayer heterostructures under external fields, and so on.

## Order parameter and BQP excitations

We applied the $GW$-BSE-SCF method to monolayer $1T'$-MoS$_2$, which has been reported as a quantum spin Hall insulator[43] with a band gap in the BI phase at the $\Lambda$ point in the BZ due to strong SOC (Fig. 2A, B). Our calculations, without any further assumptions, reveal that the order parameter $\Delta_{vc\mathbf{k}}(T)$ are nonzero (below a critical temperature $T_c$) only within a small region of the BZ centered at the $\Gamma$ point, enclosing the $\Lambda$ and $-\Lambda$ points (which are nonequivalent in the EI phase), and it is sizeable only for band indices of the two-fold degenerate valence bands ($v = 1, 2$) and conduction bands ($c = 1, 2$) nearest to the Fermi level $E_F$ (set as energy zero in Fig. 2B). In a multi-band system, $\Delta_{vc\mathbf{k}}(T)$ forms a complex-valued matrix for a given $\mathbf{k}$. For our system, it is basically a two-by-two matrix. Figure 2C shows the computed Euclidean (or Frobenius) norm of the order parameter matrix $\Delta_{vc\mathbf{k}}(T)$, denoted as $\|\Delta_{\mathbf{k}}(T)\| = \sqrt{\sum_{vc}|\Delta_{vc\mathbf{k}}(T)|^2}$, at various wavevector $\mathbf{k}$. Due to the strong electron-hole Coulomb interaction (e.g., it is much stronger than the phonon-mediated electron-electron interaction in BCS superconductivity), the maximum value of $\|\Delta_{\mathbf{k}}(T)\|$, on the order of 200 meV, is orders of magnitude larger than the typical superconducting order parameter (~1-10 meV)[44,45]. The values of $\|\Delta_{\mathbf{k}}(T)\|$ exhibit different overall temperature dependencies at different $\mathbf{k}$, but they share the same asymptotic behaviour near $T_c$, described by $\|\Delta_{\mathbf{k}}(T)\| = \Delta_{\mathbf{k}}^0\sqrt{1-(T/T_c)^{1/2}}$ as expected from a mean-field theory. Fitting the computed $\|\Delta_{\mathbf{k}}(T)\|$ near $T_c$ at the $\Lambda$ and Z point yields $\Delta_{\mathbf{k}}^0$ values of 0.53 and 0.14 eV, respectively, with a same $T_c$ of 889 K (solid lines in Fig. 2C, right panel). For $T > T_c$, $\|\Delta_{\mathbf{k}}(T)\|$ vanishes simultaneously for all $\mathbf{k}$ (Fig. 2D), with the exception at the $\Gamma$ point where it is zero at all temperatures (due to the odd parity symmetry of the order parameter; see later discussions).

The order parameter $\Delta_{vc\mathbf{k}}(T)$ determines the properties of the EI phase—in particular the BQP energies $E_n(\mathbf{k}, T)$, which enter to spectroscopic measurements associated with adding an electron to or removing an electron from the system. For monolayer $1T'$-MoS$_2$ with both a doubly degenerate valence band and conduction band near $E_F$, the BQP energies are eigenvalues of a four-by-four matrix that depends on $\Delta_{vc\mathbf{k}}(T)$ (given by equation (S39) in Supplementary Materials). As illustrated in Fig. 2E, the BQP band gaps at $T = 0$ are much larger than the band gaps of the BI phase for $\mathbf{k}$ states with nonzero $\Delta_{vc\mathbf{k}}$ (fig. S3). This increases the BQP direct band gap at the $\Lambda$ point to 0.32 eV and the BQP indirect band gap (between the $\Lambda$ and Z points) to 0.21 eV at $T = 0$, as depicted in Fig. 2F. These gaps are significantly larger than those in the BI phase, which are only 0.10 eV and 0.05 eV, respectively. Figure 2G shows the temperature-dependent DOS of the quasiparticle excitations (see Supplementary Materials section 5 'Single-particle properties of EI phase'), with the energy zero set at the computed chemical potential $\mu(T)$. In the EI phase, two prominent peaks are observed at both positive (conduction band) and negative (valence band) energies. The two larger peaks are attributed to BQP states near the Γ and Λ points, while the smaller peaks arise from BQP states near the Z point. As the temperature increases, these peaks shift in position and height, gradually reducing band gaps.

While single-particle excitation energies can be measured using techniques such as angle-resolved photoemission spectroscopy (ARPES) or inferred from tunneling experiments, these measurements do not yield a clear qualitative distinction between an EI phase from a conventional BI phase, as both are insulating. Although the BQP energies exhibit strong temperature dependence, rapid changes only occur near $T_c$, which is very high in our system. Below room temperature, these excitation energies remain nearly temperature-independent. One potential approach to facilitate experimental verification is to reduce $T_c$. This could be achieved by decreasing the electron-hole interaction strength via enhanced dielectric screening, for example, by adding a suitable substrate or considering few-layers 1$T'$-MoS$_2$ instead of a monolayer, though this would require significant materials engineering efforts to achieve a suitable $T_c$.

Here, inspired by the behaviours of some unconventional superconductors (where not only the $U(1)$ symmetry is broken), we look for an EI phase that is unconventional, breaking more than just the $U_X(1)$ symmetry. Indeed, we show that this is the case in monolayer 1$T'$-MoS$_2$, leading to the discovery of several definitive features of the EI phase that are accessible to experiments and can be used directly to identify its existence.

**Spontaneous breaking of additional symmetries**

The crystal structure of monolayer 1$T'$-MoS$_2$ possesses point group symmetry $C_{2h}$, which includes three non-trivial symmetry operations: inversion symmetry ($\hat{I}$), a mirror reflection across a plane perpendicular to the zigzag direction (defined as the $y$-axis, $\hat{m}_y$), and 180-degree rotation about the $y$-axis ($\hat{c}_{2y}$). To reveal possible breaking of any of the point group symmetries, as well as the time-reversal symmetry ($\hat{T}$), we directly examine the symmetry properties of the order parameter from our $GW$-BSE-SCF calculations. In the BI phase, both the highest valence band complex and lowest conduction band complex are doubly degenerate (owing to inversion and time-reversal symmetry). For simplicity and better physical understanding, we analyze the order parameter matrix $\Delta_{vc\mathbf{k}}(T)$ by transforming the basis from the conventional band states ($v = 1, 2$ and $c = 1, 2$) to a particular linear combination of the two states within each degenerate complex (labeled by the so-called pseudo-spin indices $\alpha$ and $\beta$) through a $\mathbf{k}$-dependent basis transformation[46], where $\alpha$ and $\beta$ each takes on two possible values: pseudo-spin up $|\uparrow\rangle$ or pseudo-spin down $|\downarrow\rangle$ (the new pseudo-spin orbitals have well-defined transformation properties; see Supplementary Materials section 6 'Manifestly covariant Bloch basis'). The order parameter matrix in the pseudo-spin basis is then denoted as $\Delta_{\alpha\beta}(\mathbf{k}, T)$, retaining the same norm as $\Delta_{vc\mathbf{k}}(T)$.

Figure 3A-D show the $\mathbf{k}$-resolved order parameter $\Delta_{\alpha\beta}(\mathbf{k})$ at $T = 0$, revealing its symmetries in our system.

In terms of usual language used in excitonic physics, the $|\uparrow\uparrow\rangle$ or $|\downarrow\downarrow\rangle$ channels correspond to electron-hole pairing of opposite spins. The participation of all pseudo-spin channels with similar amplitudes in the electron-hole pairing process, driven by the strong SOC in the system[43], suggests that any analysis based on the assumption of either pure pseudo-spin singlet or triplet pairing would be insufficient, even qualitatively, for our system.

The transformations of $\Delta_{\alpha\beta}(\mathbf{k})$ under point group and time-reversal symmetry operations are schematically illustrated in Fig. 3E-H (see Supplementary Materials section 7 'Point group transformations of order parameter' and section 8 'Time-reversal transformation of order parameter', and fig. S4). If inversion symmetry $\hat{I}$ exists, under the operation $\hat{I}$, $\Delta_{\alpha\beta}(\mathbf{k}) \Rightarrow (-1)^l \Delta_{\alpha\beta}(-\mathbf{k})$, with $l$, referred to as parity, being even. However, we find here that inversion symmetry is completely broken, as all the components of $\Delta_{\alpha\beta}(\mathbf{k})$ are not symmetric with respect to $\mathbf{k} = 0$ (the $\Gamma$ point). More directly, the difference between $\Delta_{\alpha\beta}(\mathbf{k})$ and its inversion-symmetry-connected counterpart $\Delta_{\alpha\beta}(-\mathbf{k})$ is non-zero (with a norm that is the same order as $\Delta_{\alpha\beta}(\mathbf{k})$ itself), as shown in Fig. 3I (see fig. S5 for temperature dependence). Instead of the expected even parity if the inversion symmetry were preserved, we find $\Delta_{\alpha\beta}(\mathbf{k}) = -\Delta_{\alpha\beta}(-\mathbf{k})$ (fig. S4), indicating that $l$ is odd. This odd parity in $\Delta_{\alpha\beta}(\mathbf{k})$ classifies the EI phase to a so-called $p$-wave type. The computed odd parity of $\Delta_{\alpha\beta}(\mathbf{k})$ restricts it to have one of three possible representations from group theory analysis (table S1), corresponding to those with: (1) the breaking of mirror symmetry $\hat{m}_y$ (representation $A_u$), (2) the breaking of rotation symmetry $\hat{c}_{2y}$ ($B_u$), or (3) the breaking of both symmetries ($A_u + B_u$). By comparing $\Delta_{\alpha\beta}(\mathbf{k})$ to its symmetry-connected counterpart (Fig. 3J, K), we find both $\hat{m}_y$ and $\hat{c}_{2y}$ symmetries are well broken (again at the level of the norm of $\Delta_{\alpha\beta}(\mathbf{k})$ itself). We also examined the effect of the time-reversal operator $\hat{T}$ on $\Delta_{\alpha\beta}(\mathbf{k})$, as shown in Fig. 3L, and found that the time-reversal symmetry is well preserved (fig. S6). The combination of odd parity and preserved time-reversal symmetry means that the order parameter is one of a unitary state (see Supplementary Materials section 9 'Unitarity of order parameter' and fig. S7), which is defined as having the properties $\hat{\Delta}(\mathbf{k})\hat{\Delta}^\dagger(\mathbf{k}) \propto \hat{\sigma}_0$ (at any $\mathbf{k}$ vector), where $\hat{\sigma}_0$ is the unit two-by-two matrix.

The spontaneous breaking of all point group symmetries, along with the emergence of two additional symmetries—odd parity and unitarity—in the EI phase, provides definitive predictions to distinguish the EI phase from the BI phase in our material. We have identified below some experimental measurements that can verify our theoretical predictions.

**Lifting of k-space energy degeneracy**

A direct manifestation of the symmetry breakings is typically reflected in the BQP energies $E_n(\mathbf{k})$, which are no longer degenerate at wavevectors connected by the crystal point group symmetry operations (Fig. 4A). Thus a key indicator is the lifting of $\mathbf{k}$-space degeneracy, quantified by a BQP energy shift defined to be $\delta E_n^{\hat{O}}(\mathbf{k}) = [E_n(\mathbf{k}) - E_n(\hat{O}\mathbf{k})]/2$, where $\hat{O}$ represents a point group symmetry operation ($\hat{O} = \hat{I}$, $\hat{c}_{2y}$, or $\hat{m}_y$ here). The energy difference (defined as $\delta E_n^{\hat{O}}(\mathbf{k}) - \delta E_n^{\hat{O}}(\hat{O}\mathbf{k})$) between states at $\mathbf{k}$ and $\hat{O}\mathbf{k}$ can be detected through, for example, ARPES measurements. In the BI phase, where no spontaneous symmetry breaking occurs, $\delta E_n^{\hat{O}}(\mathbf{k}) = 0$ for the two-fold degenerate valence and conduction band states in monolayer $1T'$-MoS$_2$. However, in the EI phase, our calculations show that the breaking of rotation symmetry $\hat{c}_{2y}$ ($k_x \Rightarrow -k_x$, $k_y \Rightarrow k_y$) or mirror symmetry $\hat{m}_y$ ($k_x \Rightarrow k_x$, $k_y \Rightarrow -k_y$) introduces an energy difference, with a maximum magnitude of around 4 meV at low temperatures. The defined energy shifts have opposite signs for the valence (Fig. 4B) and conduction (Fig. 4C) band states at a given $\mathbf{k}$ owing to the unitary of $\Delta_{\alpha\beta}(\mathbf{k})$ (see Supplementary Materials section 10 'BQP excitation energies with unitary order parameter'). For monolayer $1T'$-MoS$_2$, although inversion symmetry $\hat{I}$ ($k_x \Rightarrow -k_x$, $k_y \Rightarrow -k_y$) is also broken in the EI phase, the energies at opposite momenta remain degenerate, protected by the odd parity of the order parameter (fig. S8).

**Asymmetric local charge distribution in LDOS**

Real-space probes, such as scanning tunning spectroscopy (STS), which measures the LDOS, can also be used to detect symmetry breakings in 2D materials. STS operating in the constant-height mode is effective for detecting mirror symmetry $\hat{m}_y$, though it is not sensitive to inversion or rotation symmetries which would involve probing quantities at different heights, as schematically shown in Fig. 4D, E. Figure 4F displays the computed LDOS of the BI phase with energy (the tip bias) set at the first peak in the DOS of the valence bands and evaluated at the S atom plane. At this energy, the LDOS is symmetric with respect to the mirror plane, as highlighted by the two line cuts perpendicular to the mirror plane. In contrast, in the EI phase at $T = 0$, the LDOS along the same line cuts is asymmetric (Fig. 4G) due to the breaking of mirror symmetry. Additional LDOS data for other bias voltage positions are presented in fig. S9.

**Emergent k-dependent electron spin polarization**

More importantly, while the above quasiparticle energy and LDOS signatures from spontaneous symmetry breakings are measurable and interesting, our calculations lead to the discovery of a novel and prominent $p$-wave spin textures emergent in the EI phase of monolayer $1T'$-MoS$_2$. In the BI phase, the combined presence of inversion and time-reversal symmetries ensures Kramers spin degeneracy of the band states, precluding any net electron-spin polarization at a given $\mathbf{k}$ point for states in the doubly degenerate valence or conduction bands (i.e., the contributions of the two states in the degenerate complex sum to zero). In the EI phase, the breaking of inversion symmetry lifts the Kramers spin degeneracy; nevertheless, each band remains doubly energy degenerate due to unitary ordering (see Supplementary Materials section 10 'BQP excitation energies with unitary order parameter'). The EI phase thus permits a $\mathbf{k}$-dependent finite net spin polarization for a given degenerate complex, characterized by a reversal of the spin orientation at wavevectors connected by time-reversal and those between the valence and conduction band states at the same $\mathbf{k}$, as schematically depicted in Fig. 4H. While the lifting of Kramers spin degeneracy in materials with broken inversion symmetry tends to be weak due to its origins in relativistic SOC[47,48], the stark transition of the ground state from even parity in the BI phase to odd parity in the EI phase means a complete break from having inversion symmetry. This significantly accentuates the magnitude of the spin polarization.

Focusing specifically on the $\Lambda$ point (with $k_x = 0$, $k_y \neq 0$), where the largest value of the order parameter occurs, we anticipate a pronounced spin polarization of the bands there. Also, with $\|\hat{\Delta}(\mathbf{k}) - \hat{c}_{2y}\hat{\Delta}(\mathbf{k})\hat{c}_{2y}^{-1}\|$ at $\Lambda$ being small (see fig. S5), it suggests a dominance of the $S_y(\mathbf{k})$ component of the electron spin polarization (as defined by equation (S24); see Supplementary Materials section 5 'Single-particle properties of EI phase') over the $S_x(\mathbf{k})$ and $S_z(\mathbf{k})$ components, since if this symmetry exists they are constrained by the transformations $S_x \Rightarrow -S_x$, $S_y \Rightarrow S_y$, $S_z \Rightarrow -S_z$ under $\hat{c}_{2y}$. As shown in Fig. 4I, the $S_y(\mathbf{k})$ component for the valence band (i.e., the sum of the expectation values of the individual states in the degenerate complex) at $\Lambda$ is gigantic, approximately 1.8 $\mu_B$ per pairs of states at $T = 0$ in the EI phase, diminishing to zero as the temperature approaches $T_c$. The $\mathbf{k}$-resolved distribution of $S_x(\mathbf{k})$, $S_y(\mathbf{k})$, and $S_z(\mathbf{k})$ at $T = 0$, depicted in Fig. 4J-L, show that all spin components achieve finite values near the $\Lambda$ points. $S_x(\mathbf{k})$ (Fig. 4J) and $S_z(\mathbf{k})$ (Fig. 4L) however manifest very minor values with maximum magnitudes of 0.07 and 0.04 $\mu_B$, respectively. The patterns of these electron spin polarizations in momentum space exhibit an apparent preservation of $\hat{c}_{2y}$ symmetry (although other physical quantities do not). Intriguingly, the behaviour of these electron spin polarization components deviates significantly from the transformations under $\hat{m}_y$ operation ($S_x \Rightarrow -S_x$, $S_y \Rightarrow S_y$, $S_z \Rightarrow -S_z$): instead of the expected results of $S_x(\mathbf{k})$ and $S_z(\mathbf{k})$ inverting their signs and $S_y(\mathbf{k})$ remining unchanged, the opposite responses are observed in the computed results.

The spin magnetic structure of the EI phase in our system thus cannot be straightforwardly classified. Conventional antiferromagnetism typically lacks distinct spin texture in momentum space, while ferromagnetism involves the breaking of time-reversal symmetry. Recent studies[49,50] have identified an unconventional magnetic phase known as altermagnetic, characterized by opposite-spin oriented sublattices connected by a rotation transformation. Although the $\mathbf{k}$-dependent spin textures in the EI phase here somewhat resemble those observed in altermagnetic materials, it is not accurate to categorize the EI phase as altermagnetic. This distinction arises because the spin magnetic moments in the EI phase result from long-range Coulomb interaction as opposed to being localized at atomic positions, and the real-space magnetization is zero everywhere (fig. S10). This asymmetric and $\mathbf{k}$-dependent spin texture, particularly evident in the prominent $S_y(\mathbf{k})$ component with two lobes of opposite signs (Fig. 4K), exemplifies what is often referred to as $p$-wave magnetism—a phenomenon hypothesized and pursued[51]. We emphasize that this $p$-wave spin texture stands out as the most compelling evidence of the EI phase in our study of monolayer $1T'$-MoS$_2$, and it should be detectable through spin-resolved and angle-resolved photoemission spectroscopy (SARPES).

**Conclusion and Outlook**

We have presented a theoretical framework for the parameter-free ab initio calculations of the order parameter, accompanied by detailed symmetry analyses, to study EIs and their properties. We show that monolayer $1T'$-MoS$_2$ is an unconventional EI, with telltale spectroscopic signatures distinguishable from those of the BI phase. These signatures can be robustly measured using **k**-space, real-space, and spin-resolved spectroscopies. Another crucial discovery from our study is a $p$-wave **k**-dependent spin texture of the BQP states in the EI phase, suggesting a new form of spin ordering in **k** space. These findings add to the current interest in exotic quantum phases with complex behaviours. Our research also opens several pathways for further exploration. Our study focused on electron-hole pairs with zero center-of-mass wavevector $\mathbf{Q} = 0$. To achieve a more comprehensive understanding of EI phases, future efforts should extend to possible finite **Q** pairings which may be important in some other materials. While this extension will significantly increase computational demands, it is essential toward a more complete exploration of EI phenomena. The spontaneous breaking of crystal translation symmetry in the EI phase with finite **Q** pairings could lead to new phenomena related to charge and spin density wave behaviours.

**Acknowledgements:** This work was primarily supported by the National Science Foundation under Grant No. DMR-2325410, which provided the development of the ab initio methods for computing the order parameter and the group theoretical and symmetry analyses of the properties of EI phases. Additional support was provided by the Center for Computational Study of Excited-State Phenomena in Energy Materials (C2SEPEM) at LBNL, funded by the U.S. Department of Energy, Office of Science, Basic Energy Sciences, Materials Sciences and Engineering Division under Contract No. DE-AC02-05CH11231, as part of the Computational Materials Sciences Program, which facilitated the development of advanced codes and simulations as well as provided the *GW* and *GW*-BSE calculations of electron-hole interactions and exciton properties. Computational resources were provided by National Energy Research Scientific Computing Center (NERSC), supported by the Office of Science of the U.S. Department of Energy under Contract No. DE-AC02-05CH11231; Stampede2 at the Texas Advanced Computing Center (TACC), the University of Texas at Austin, through Extreme Science and Engineering Discovery Environment (XSEDE), supported by the National Science Foundation under Grant No. ACI-1053575; and Frontera at TACC, supported by the National Science Foundation under Grant No. OAC-1818253. We thank Xiaoxun Gong, Weichen Tang, Dung-Hai Lee, and Michael Zaletel for the fruitful discussions.

**Competing interests:** The authors declare no competing interests.

**Data and materials availability:** All data supporting this study are provided in the main text or Methods. The calculations presented in the paper were carried out using publicly available electronic structure codes as described in Methods. Our findings can be fully reproduced by the use of these codes and following the procedure outlined in the paper.

**Correspondence:** Correspondence and requests for original materials should be addressed to [sglouie@berkeley.edu](sglouie@berkeley.edu) (S.G.L.).

**Supplementary Materials:** Supplementary Text Sections 1 to 10; Figs. S1 to S10; References (52-60)

**Main figures and captions**

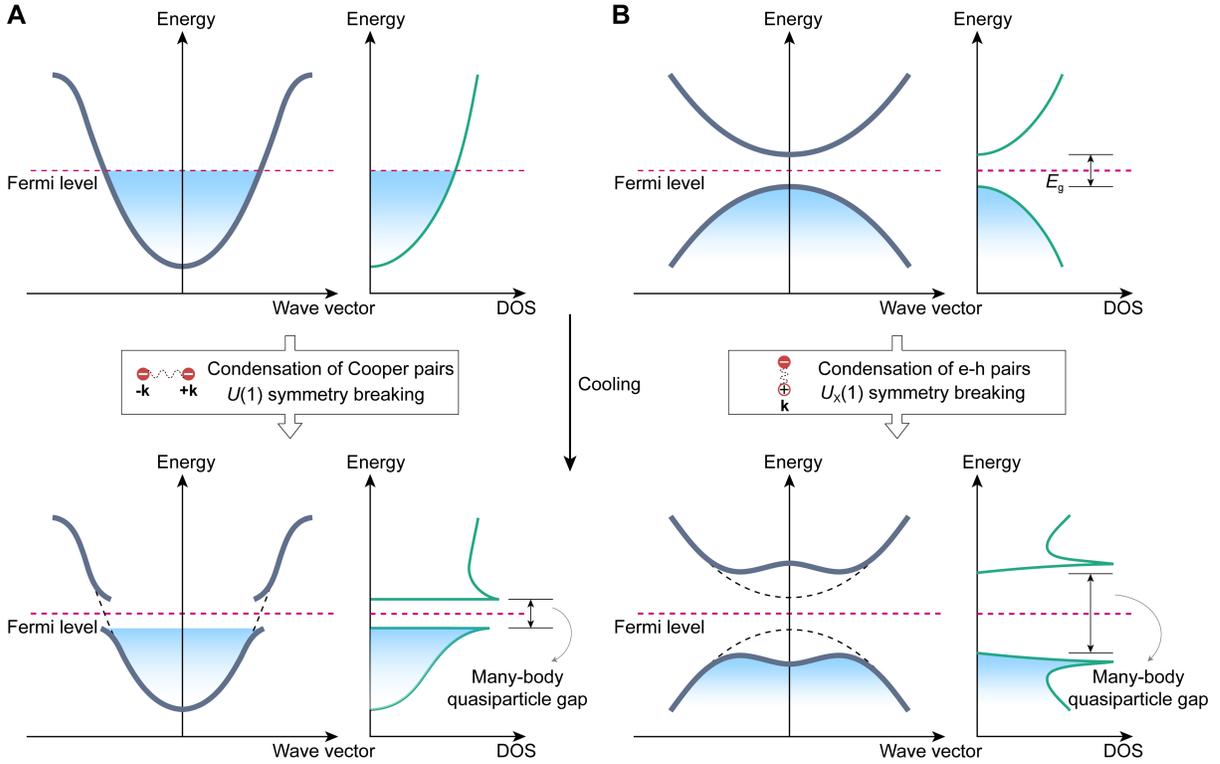

**Fig. 1: Schematics of Bogoliubov quasiparticle excitations in superconductors and excitonic insulators.** **A,** Upper panels: Schematic band structure (left) and quasiparticle density of states (DOS) (right) of a metal, with the Fermi level highlighted by red dashed lines. Lower panels: Transition from the normal metallic phase into a superconducting phase at low temperature. In the superconducting phase, electrons with opposite momenta pair up and condense, forming Cooper pairs and breaking $U(1)$ symmetry. This transition typically introduces a many-body superconducting quasiparticle gap that significantly alters both the Bogoliubov quasiparticle (BQP) band structure (left) and the DOS (right). **B**, Upper panels: Schematic band structure (left) and DOS (right) of a conventional band insulator (BI), characterized by a band gap $E_g$. Lower panels: Below a critical temperature, electron-hole (e-h) pairs, or excitons, spontaneously form and condense. This condensation breaks $U_X(1)$ symmetry, resulting in a many-body insulating state known as an excitonic insulator (EI) with a new quasiparticle band structure (left) and DOS (right).

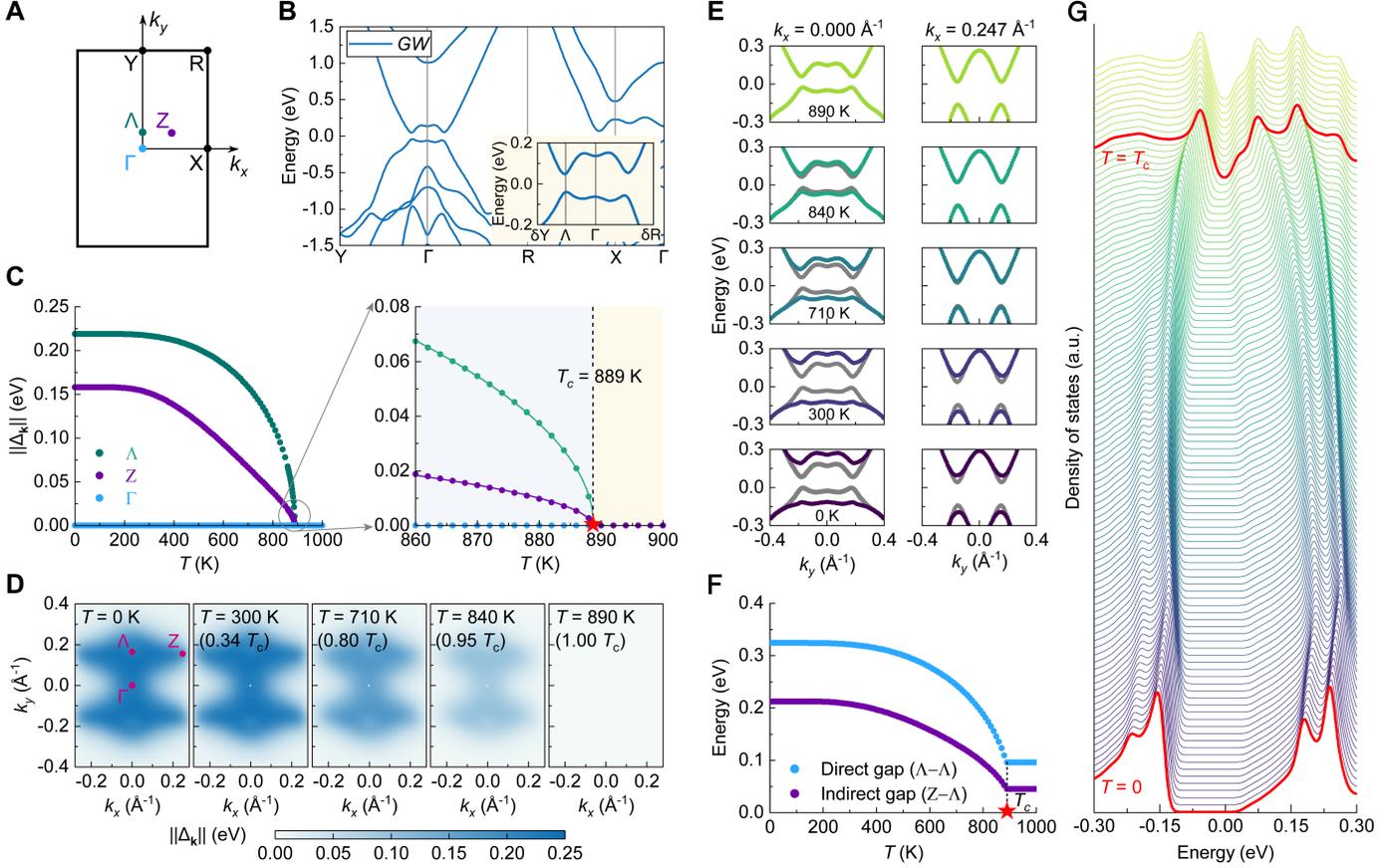

**Fig. 2: Order parameter and temperature-dependent Bogoliubov quasiparticle excitations in monolayer $1T'$-MoS$_2$. A, B**, The first Brillouin zone (BZ) (**A**) and *GW* band structures for conventional BI phase (**B**) calculated with spin-orbital coupling (SOC). The valence band maximum and conduction band minimum are located at $\Lambda$ and $Z$ points, respectively. The minimum direct band gap is located at the $\Lambda$ point (inset of **B**). **C**, Left: temperature dependence of the Frobenius norm of the order parameter matrix, $\|\Delta_{\mathbf{k}}(T)\| = \sqrt{\sum_{vc}|\Delta_{vc\mathbf{k}}(T)|^2}$, at the $\Lambda$, $Z$, and $\Gamma$ points. Right: $\|\Delta_{\mathbf{k}}(T)\|$ at temperatures close to the transition temperature $T_c$. The solid curves at the $\Lambda$ and $Z$ points are results of best fits to the form of $\|\Delta_{\mathbf{k}}(T)\| = \Delta_{\mathbf{k}}^0 \sqrt{1-(T/T_c)^{0.5}}$, yielding the same $T_c = 889$ K (red star). **D**, Momentum-space distribution of $\|\Delta_{\mathbf{k}}(T)\|$ at different temperatures. Outside of the plotted regions of the BZ, $\|\Delta_{\mathbf{k}}(T)\|$ is smaller than 6 meV at $T = 0$. **E**, Temperature dependence of the calculated BQP excitation energies $E_n(\mathbf{k})$ (different bright colors) along the $k_y$ path with fixed $k_x = 0$ (left panels), and $k_x = 0.247$ Å$^{-1}$ (right panels). The band energies $\varepsilon_n(\mathbf{k})$ in the BI phase are shown as reference (gray dots). In each panel, the chemical potential $\mu(T)$ is set as energy zero. **F**, Temperature dependence of the direct band gap at the $\Lambda$ point, and the indirect band gap between the $Z$ and $\Lambda$ points. **G**, DOS of BQPs ranging from $T = 0$ K to 1,000 K with 10 K interval, calculated using 15 meV Lorentzian smearing. The DOS at $T = 0$ and $T = T_c$ are heighted by red colors. The data points in (**G**) are vertically shifted for easy comparison.

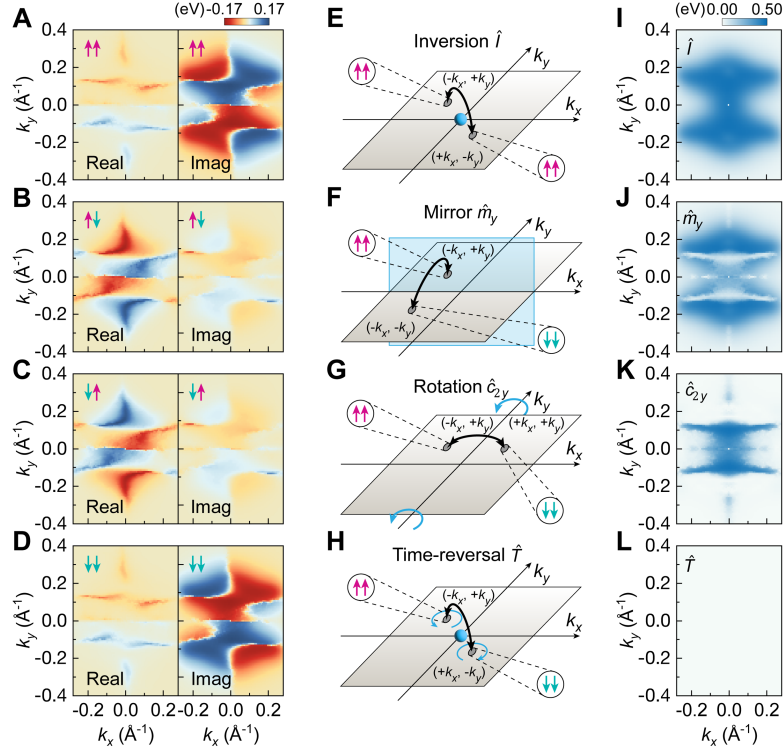

**Fig. 3: Symmetry transformations of the order parameter in monolayer $1T'$-MoS$_2$. A-D**, The momentum-space distributions of the real part (left panels) and imaginary part (right panels) of the $|\uparrow\uparrow\rangle$ (A), $|\uparrow\downarrow\rangle$ (B), $|\downarrow\uparrow\rangle$ (C), and $|\downarrow\downarrow\rangle$ (D) components of the order parameter $\Delta_{\alpha\beta}(\mathbf{k})$ in the pseudo-spin basis. All pseudo-spin channels show significant contributions, highlighting the mixing of pseudo-spin triplet and singlet pairing. **E-H**, Schematics of the operation of inversion $\hat{I}$ (E), a mirror reflection across a plane perpendicular to the $y$-axis $\hat{m}_y$ (F), a 180-degree rotation around the $y$-axis $\hat{c}_{2y}$ (G), and time-reversal $\hat{T}$ (H). Each operation affects both the wavevector $\mathbf{k}$ and the pseudo-spin indices due to the strong SOC. **I-L**, The Frobenius norm of the *difference* between the original order parameter matrix and its symmetry-transformed counterpart, for the inversion operation $\Delta_{\alpha\beta}(\mathbf{k}) - \hat{I}\Delta_{\alpha\beta}(\mathbf{k})\hat{I}^{-1}$ (I), mirror operation $\Delta_{\alpha\beta}(\mathbf{k}) - \hat{m}_y\Delta_{\alpha\beta}(\mathbf{k})\hat{m}_y^{-1}$ (J), rotation operation $\Delta_{\alpha\beta}(\mathbf{k}) - \hat{c}_{2y}\Delta_{\alpha\beta}(\mathbf{k})\hat{c}_{2y}^{-1}$ (K), and time-reversal operation $\Delta_{\alpha\beta}(\mathbf{k}) - \hat{T}\Delta_{\alpha\beta}(\mathbf{k})\hat{T}^{-1}$ (L). The computed large Frobenius norms in (I-K) indicate that $\hat{I}$, $\hat{m}_y$, and $\hat{c}_{2y}$ symmetries are broken (to order of unity) in the EI phase, while $\hat{T}$ symmetry remains preserved within our numerical accuracy.

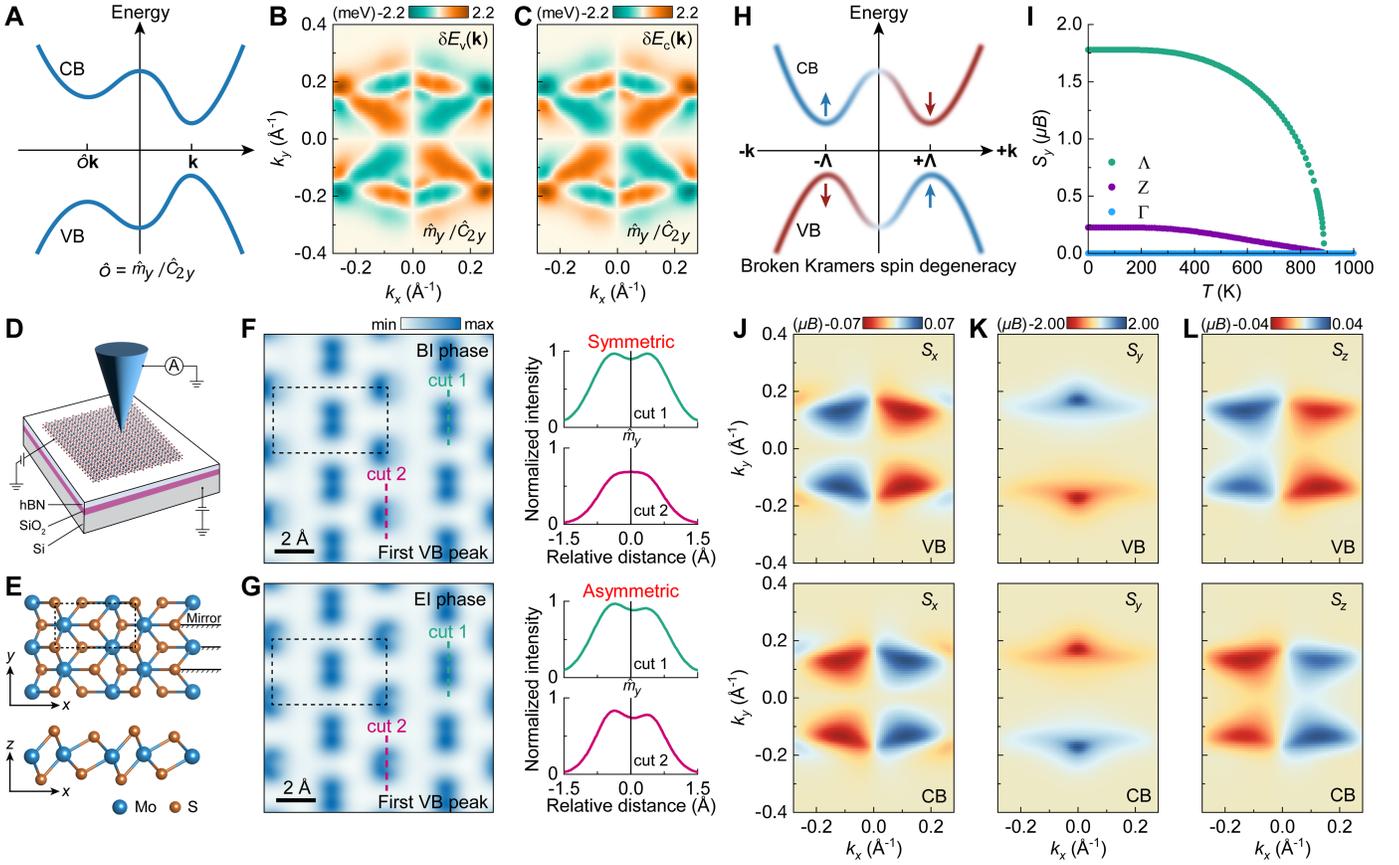

**Fig. 4: Spectroscopic signatures from breaking of symmetries in EI phase of monolayer 1T′-MoS$_2$. A,** Schematic band structure along a direction for states **k** and the corresponding transformed $\hat{O}\mathbf{k}$ direction, where $\hat{O}$ denotes either the mirror $\hat{m}_y$ or rotation $\hat{C}_{2y}$ operations. BQP excitation energies at symmetry-connected wavevectors are now different due to spontaneous symmetry breaking in the EI phase. **B, C,** Calculated shifts in BQP excitation energy, $\delta E_n^{\hat{O}}(\mathbf{k}) = [E_n(\mathbf{k}) - E_n(\hat{O}\mathbf{k})]/2$, for the valence band states (**B**) and the conduction band states (**C**). The results are found to be identical for $\hat{O} = \hat{m}_y$ and $\hat{O} = \hat{C}_{2y}$. **D,** Schematic setup of a scanning tunning spectroscopy (STS) experiment. **E,** Top view (upper panel) of crystal structure of monolayer 1T′-MoS$_2$, with the unit cell outlined by black dashed lines and mirror symmetry planes indicated, and side view (lower panel). **F, G,** Computed local density of states (LDOS) map evaluated on a plane at the S atomic layer in the BI phase (**F**) and the EI phase (**G**), with the tip bias voltage set to the first peak in the DOS of the valence band complex (see Fig. 2**G**). The unit cell is marked by black dashed lines. Left panels are two-dimensional maps and right panels are line plots for the indicated two line cuts. **H,** Schematic of the electron spin texture of the band states along $k_y$ ($k_x = 0$), including the two Λ points. Up and down arrows indicate negative and positive spin expectation values $S_y(\mathbf{k})$ of the electron. **I,** Computed temperature dependence of the $S_y(\mathbf{k})$ component of the electron spin expectation value at selected **k** points. **J-L,** The electron spin expectation values along the $x$- (**J**), $y$- (**K**), and $z$- (**L**) directions, with the upper and

19 / 20

bottom panels corresponding to valence and conduction band states, respectively. Note that the color scale is different for (**K**) as compared to those for (**J**) and (**L**).